\documentclass[preprint,aps,nofootinbib]{revtex4}
\usepackage{}
\usepackage{graphicx}%Include figure files
\usepackage{amsmath}
\usepackage{amsfonts}
\usepackage{amssymb}
\usepackage{color}%
\usepackage{dcolumn}% Align table columns on decimal point
\usepackage{indentfirst}
\providecommand{\U}[1]{\protect\rule{.1in}{.1in}}

\definecolor{lightgray}{rgb}{.7,.7,.7}

\definecolor{red}{rgb}{1,0,0}

\definecolor{blue}{rgb}{0,0,1}

\definecolor{purple}{rgb}{0.6,0.1,0.7}

\newcommand{\f}{\begin{equation}}
\newcommand{\ff}{\end{equation}}
\newcommand{\fa}{\begin{eqnarray}}
\newcommand{\ffa}{\end{eqnarray}}

\begin{document}
\title{Note on the emergence of cosmic space in modified gravities}
\author{Yi Ling $^{1,2,3}$}
\email{lingy@ihep.ac.cn}
\author{Wen-Jian Pan $^{2,1}$}
\email{wjpan_zhgkxy@163.com} \affiliation{$^1$Institute of High
Energy Physics, Chinese Academy of Sciences, Beijing 100049,
China\\ $^2$ Center for Relativistic Astrophysics and High Energy
Physics, Department of Physics, Nanchang University, 330031, China
\\ $^3$ State Key Laboratory of Theoretical Physics, Institute of Theoretical Physics, Chinese Academy of Sciences, Beijing 100190}

\begin{abstract}
Intrigued by the holographic principle, Padmanabhan recently
proposed a novel idea, saying that our cosmic space is emergent as
cosmic time progresses. In particular, the expansion rate of the
Universe is related to the difference between the surface degrees
of freedom on the holographic horizon and the bulk degrees of
freedom inside. In this note, we generalize this interesting
paradigm to brane world, scalar-tensor gravity, and f(R) theory, respectively,
and find that in the cosmological setting the Friedmann equations
can be successfully derived.
\end{abstract}
\maketitle

\section{Introduction}
The tight link between a gravitational system and a
thermodynamical system was discovered in
1970s\cite{Bardeen:1973gs}. A thermodynamical description of the
Einstein equation was originally proposed by Jacobson
\cite{Jacobson:1995ab}. In this scenario the Einstein equation
becomes an equation of state and can be derived from a fundamental
thermodynamical relation, namely Clausius relation $\delta Q =
TdS$, which connects heat, entropy, and temperature for all the
local Rindler causal horizons. Recently, this scenario has been
demonstrated in many gravity theories and cosmological
models\cite{Padmanabhan:2002sha,Kothawala:2007em,Cai:2005ra,Akbar:2006mq,Cai:2008mh,Eling:2006aw,Sheykhi:2007gi,Ge:2007yu,
Gong:2007md,Wu:2007se,Ling:2009wj,Guedens:2011dy,Sharif:2012zzd}.
In this context, the dynamical equations of the gravitational
field can be derived from the Clausius relation. Furthermore, the
viewpoint that gravity is not a fundamental interaction has been
developed in the paper by Verlinde \cite{Verlinde:2010hp}. Gravity
is explained as an entropic force that is caused by the change of
the information associated with the position of matter. In
\cite{Padmanabhan:2009kr}, Padmanabhan also argued that the
equipartition law of energy for the horizon degrees of freedom can
lead to Newton's law of gravity through the thermodynamical
relation $S={E\over2T}$, where $E$ is the active gravitational
mass producing the gravitational acceleration in spacetime
\cite{Padmanabhan:2003pk}.

Very recently, Padmanabhan presented a novel way to view the
cosmic space as an emergent phenomenon in a cosmological
setting\cite{Padmanabhan:2012ik}. The main idea is attributing the
expansion of our Universe to the difference between the surface
degrees of freedom on the holographic horizon and the bulk degrees of
freedom in bulk. In this paradigm the dynamical equation of a
Friedmann-Robertson-Walker (FRW)
universe can be successfully derived. After that, this setup
has been generalized to the cosmology in Gauss-Bonnet gravity and
more general Lovelock gravity in\cite{Cai:2012ip,Yang:2012wn}.
  Further discussions by Padmanabhan on its applications in the cosmology
   can be found in\cite{Padmanabhan:2012gx}.

We may briefly summarize the idea proposed by Padmanabhan as
follows. For a pure de Sitter universe with Hubble parameter $H$,
the holographic principle can be described by the relation
 \begin{equation}
N_{sur}=N_{bulk},
\end{equation}
where $N_{sur}$ denotes the number of the degrees of freedom on
the holographic screen with Hubble radius $1/H$, namely
$N_{sur}={4\pi\over L^2_pH^2}$ with $L_p^2=G$, while
$N_{bulk}={2|E|\over T}$ is the number of the degrees of freedom
in bulk, where $|E|=|\mathcal{M}|=|\rho+3p|V$ is the Komar mass.
The horizon temperature is determined by $T={H\over2\pi}$. The
above equation is the so-called holographic equipartition. Since
the real world is not purely but asymptotically de Sitter,
one may propose that the expansion rate of the cosmic volume is
related to the difference of these two degrees of freedom as
\begin{equation}
{dV\over dt}=L^2_p(N_{sur}-N_{bulk})\label{spem}.
\end{equation}
As shown in \cite{Padmanabhan:2012ik}, Eq. (\ref{spem}) shows that
it is very necessary for the existence of the cosmological constant to
drive the expansion of the Universe toward holographic
equipartition.\footnote{We remark that using the Hubble
horizon, one can find that Eq.(2) is consistent with the
Raychaudhuri's equation in the FRW universe. We thank the anonymous
referee for raising this issue with us.} Substituting the cosmic
volume $V={4\pi\over3H^3}$ and the number of the degrees of
freedom into the above equation, we obtain the standard dynamical
Friedmann equation in general relativity,
 \begin{equation}
{\ddot{a}\over{a}}=-{4\pi G\over3}(\rho+3p)\label{dy0}.
\end{equation}
In addition, making use of the continuity equation,
\begin{equation}
\dot{\rho}+3H(\rho+p)=0,
\end{equation}
and integrating Eq.(\ref{dy0}), one can obtain the Friedmann equation,
\begin{equation}
H^2={8\pi G\over3}\rho,
\end{equation}
where the integration constant has been set to zero.

In this paper, we will examine this interesting proposal in
other gravitational theories. We intend to generalize this
paradigm to the brane world, scalar-tensor gravity and f(R)
gravity. The key issue is how to find the correct result for
the Komar mass in these modified gravity theories since it plays
an essential role in this approach. Our strategy is to introduce a
total effective energy-momentum tensor such that the equations of
motion have the same form as the one in standard general
relativity. As a result we can derive the Komar mass in a similar
way as that in general relativity.

Our paper is organized as follows. In Sec. II, we treat the
expansion of the cosmic space as an emergent process and derive
the Friedmann equations in the context of the brane world, while in
Secs. III and IV, we intend to derive the Friedmann equations in
scalar-tensor theory and f(R) gravity, respectively.
 Our conclusion and discussion are given in the last section.
 We shall set the constants $\hbar=c=k_B=1$ in this paper.

\section{Dynamical Friedmann equations as an emergence of the cosmic space in the brane world}
In this section, we consider how this paradigm can deduce the
equation of motion of our Universe in the context of the brane
world. First of all, we introduce the metric of the background as
\begin{equation}
ds^2=-dt^2+a^2(t)[{dr^2\over 1-kr^2}+r^2(d\theta^2+\sin^2\theta d\phi^2)]
\end{equation}
where $k=1, 0, -1$ corresponds to a closed, a flat, and an open
universe, respectively. Without loss of generality, we may only
consider the spatially flat FRW universe without cosmological
constant embedded in a Randall-Sundrum model. In a brane world,
the key difference is to determine the Komar mass in terms of $T_{ab}$.
In Ref.\cite{Ling:2010zc} the Komar mass for a brane world
has been derived as
\begin{eqnarray}
\mathcal{M} &=&\int_{\upsilon}dV(\rho+3P+{2\rho^2\over\lambda}+{3\rho P\over\lambda})\nonumber\\
&=&{4\pi\over3H^3}(\rho+3P+{2\rho^2\over\lambda}+{3\rho
P\over\lambda}),\label{KOM}
\end{eqnarray}
where we have identified the Hubble horizon as the radius of the
holographic screen and $\lambda$ is the brane tension which is
tuned with the five-dimensional cosmological constant by
$\lambda\sim-{\Lambda_5\over8\pi G_4}$. Furthermore, it is worthy
noting that we have used the total effective energy-momentum tensor for
the brane world. Substituting Eq.(\ref{KOM})into Eq.(\ref{spem}), we
have
\begin{eqnarray}
{dV\over dt}&=&L^2_p({A\over G}-{-2\mathcal{M}\over T})\nonumber\\
&=&{4\pi\over
H^2}+{16\pi^2G\over3H^4}(\rho+3P+{2\rho^2\over\lambda}+{3\rho
P\over\lambda}).
\end{eqnarray}
Simplifying the above equation leads to the standard dynamical
Friedmann equation in brane cosmology,
 \begin{equation}
{\ddot{a}\over{a}}=-{4\pi G\over3}(\rho+3P+{2\rho^2\over\lambda}+{3\rho P\over\lambda})\label{dy1}.
\end{equation}
Moreover, using the continuity equation $\dot{\rho}+3H(\rho+P)=0$,
one can easily derive the other Friedmann equation,
\begin{equation}
H^2={8\pi G\over3}\rho(1+{\rho\over2\lambda})\label{kem1}.
\end{equation}

\section{Dynamical Friedmann equations as an emergence of the cosmic space in scalar-tensor gravity }
Now we in turn consider the emergence of the cosmic space in
scalar-tensor gravity. The equation of motion in scalar-tensor
gravity can be written as \cite{Akbar:2006er}
\begin{eqnarray}
G_{\mu\nu}&=&R_{\mu\nu}-{1\over2}Rg_{\mu\nu}\nonumber\\
&=&{8\pi G\over f(\phi)}[\nabla_{\mu}\phi\nabla_{\nu}\phi-{1\over2}g_{\mu\nu}(\nabla\phi)^2
-g_{\mu\nu}V(\phi)-g_{\mu\nu}\nabla^2f(\phi)+\nabla_{\mu}\nabla_{\nu}f(\phi)+T^{(m)}_{\mu\nu}],\label{Efm}
\end{eqnarray}
and
\begin{equation}
\nabla^2\phi-V^{\prime}(\phi)+{1\over2}f^{\prime}(\phi)R=0
\end{equation}
where $G_{\mu\nu}$ and $T^{(m)}_{\mu\nu}$ denote the Einstein
tensor and the energy-momentum tensor of matter, respectively,
while $R$ is the Ricci curvature scalar. $f(\phi)$ is a generic
function of a scalar field $\phi$. Here we only consider the flat
universe. For convenience, we can define an energy-momentum tensor
for the scalar field,
\begin{equation}
T^{(\phi)}_{\mu\nu}=\nabla_{\mu}\phi\nabla_{\nu}\phi-{1\over2}g_{\mu\nu}(\nabla\phi)^2
-g_{\mu\nu}V(\phi)-g_{\mu\nu}\nabla^2f(\phi)+\nabla_{\mu}\nabla_{\nu}f(\phi).
\end{equation}
Moreover, we assume the energy-momentum tensor $T^{(\phi)}_{\mu\nu}$ has
the same form as that of a perfect fluid,
\begin{equation}
T_{\mu\nu}=(\rho+P)u_{\mu}u_{\nu}+Pg_{\mu\nu}.
\end{equation}
It is worthwhile to stress that such an assumption is compatible
with the hypothesis of homogeneity and isotropy of the Unverse.
Then, correspondingly, we can find the energy density as
well as the pressure of the scalar field as
\begin{eqnarray}
\rho^{(\phi)}&=&{1\over2}(\dot{\phi})^2+V-3H\dot{f},\label{cmd}\\
P^{(\phi)}&=&{1\over2}(\dot{\phi})^2-V+2H\dot{f}+\ddot{f}.\label{cpq}
\end{eqnarray}
Taking the ordinary matter into account, we can further define a
total effective energy-momentum tensor as
\begin{eqnarray}
T^{(t)}_{\mu\nu}&=&[{\rho^{(m)}+P^{(m)}+\rho^{(\phi)}+P^{(\phi)}\over f(\phi)}]u_{\mu}u_{\nu}+[{P^{(m)} +P^{(\phi)}\over f(\phi)}]g_{\mu\nu}\nonumber\\
&=&[{\rho^{(m)}+P^{(m)}+(\dot{\phi})^2-H\dot{f}+\ddot{f}\over f(\phi)}]u_{\mu}u_{\nu}
+[{P^{(m)}+{1\over2}(\dot{\phi})^2-V+2H\dot{f}+\ddot{f}\over f(\phi)}]g_{\mu\nu}.
\end{eqnarray}
As a result, the equation of motion in scalar-tensor theory can be
rewritten as a compact form,
\begin{equation}
G_{\mu\nu}=8\pi GT^{(t)}_{\mu\nu}\label{Ein}.
\end{equation}
Now we consider deriving the dynamic Friedmann equations
in scalar-tensor theory from an emergent point of view. Due to
the above equation, we should still adopt the assumptions
$N_{sur}={A\over G}={4\pi\over GH^2}$ and $T={H\over2\pi}$
in scalar-tensor theory.\footnote{Theoretically, one expects that
such assumptions should be supported by the thermodynamics
of black holes in scalar-tensor gravity. Such subtleties
has previously been analyzed in \cite{Akbar:2006er}. }
Next, the key step is to find out the effective Komar mass in scalar-tensor
theory. Since the equation of motion for scalar-tensor gravity has the same form as
the one in standard general relativity, we argue that the Komar
mass in scalar-tensor cosmology can still be evaluated as
\begin{eqnarray}
\mathcal{M} &=&2\int_{\upsilon}dV[T^{(t)}_{\mu\nu}-{1\over2}T^{(t)}g_{\mu\nu}]u^{\mu}u^{\nu}\nonumber\\
&=&V[{\rho^{(m)}+3P^{(m)}\over f(\phi)}+{2(\dot{\phi})^2-2V+3H\dot{f}+3\ddot{f}\over f(\phi)}].\label{KOM2}
\end{eqnarray}
As a consequence, the expansion rate of the Universe in scalar-tensor
cosmology can be obtained from Eq.(\ref{spem}) as
\begin{equation}
{\ddot{a}\over{a}}=-{4\pi G\over3}({\rho^{(m)}+3P^{(m)}\over f(\phi)}+{2(\dot{\phi})^2-2V+3H\dot{f}+3\ddot{f}\over f(\phi)})\label{dy2}.
\end{equation}
Using the effective energy density in Eq.(\ref{cmd}) and pressure
density in Eq.(\ref{cpq}), we can write this equation in the
familiar form,
\begin{equation}
{\ddot{a}\over{a}}=-{4\pi G\over3}({\rho^{(m)}+3P^{(m)}\over f(\phi)}+{\rho^{(\phi)}+3P^{(\phi)}\over f(\phi)}).
\end{equation}
Moreover, due to the Bianchi identity, $\nabla^{\mu}G_{\mu\nu}=0$,
we have the conservational relation of the total effective
energy-momentum tensor $\nabla^{\mu}T^{(t)}_{\mu\nu}=0$, which
gives the continuity equation as
\begin{equation}
{d\over dt}({\rho^{(\phi)}+\rho^{(m)}\over f(\phi)})+3H({\rho^{(m)}+P^{(m)}+\rho^{(\phi)}+P^{(\phi)}\over f(\phi)})=0.
\end{equation}
Making use of the above continuity equation and integrating
Eq.(\ref{dy2}), we can finally obtain
 the other Friedmann equation as
\begin{equation}
H^2={8\pi G\over3}[{\rho^{(m)}+\rho^{(\phi)}\over f(\phi)}].
\end{equation}
Here we have also set the integration constant to vanish.
\section{Dynamical Friedmann equations as an emergence of the cosmic space in $f(R)$ theory }
In the last section, using the same tactics, we consider the
emergence of the cosmic space in $f(R)$ theory concisely. In the
literature
\cite{Akbar:2006er,Capozziello:2005ku,Nojiri:2006ri,Nojiri:2010wj,DeFelice:2010aj,Sotiriou:2008rp,Jaime:2012gc,Bamba:2011jq},
 the equation of motion in f(R) gravity is given by
\begin{eqnarray}
G_{\mu\nu}&=&R_{\mu\nu}-{1\over2}Rg_{\mu\nu}\nonumber\\
&=&8\pi G[{T^{(m)}_{\mu\nu}\over f'(R)}+{f(R)g_{\mu\nu}\over16\pi Gf'(R)}-{Rf'(R)g_{\mu\nu}\over16\pi Gf'(R)}+{1\over8\pi Gf'(R)}(\nabla_{\mu}\nabla_{\nu}-g_{\mu\nu}\Box)f'(R)],\label{Efm}
\end{eqnarray}
where $T^{(m)}_{\mu\nu}$ is the energy-momentum tensor of matter
and $f'\equiv{df(R)\over dR}$. In the cosmological setting, f(R) is
an arbitrary function of Ricci scalar curvature $R$ with
$R=6(\dot{H}+2H^2+{k\over a^2})$. Here we only consider the flat
universe with $k=0$. For simplicity, we may define a
total effective energy-momentum tensor for both
 ordinary matter and effective curvature
fluid, which reads as \cite{Akbar:2006er}
\begin{eqnarray}
T^{(t)}_{\mu\nu}&=&[{\rho^{(m)}\over f'}+{P^{(m)}\over f'}+\rho^{(c)}+P^{(c)}]u_{\mu}u_{\nu}+[{P^{(m)}\over f'}+P^{(c)}]g_{\mu\nu}
\end{eqnarray}
where $\rho^{(c)}={R f'-f-6H\dot{R}f^{\prime\prime}\over16\pi G
f'}$ and
$P^{(c)}={\dot{R}^2f^{\prime\prime\prime}+2H\dot{R}f^{\prime\prime}+\ddot{R}f^{\prime\prime}-{1\over2}(R
f'-f)\over8\pi Gf^{\prime}}$. As a result, the equation of motion
in f(R) theory can also be written as a compact form,
as in Eq.(\ref{Ein}). In a parallel way, we obtain the Komar mass in
f(R) theory and then plug it into Eq.(\ref{spem}), such that the
dynamical equation in $f(R)$ cosmology can be obtained as
\begin{equation}
{\ddot{a}\over{a}}=-{4\pi G\over3}({\rho^{(m)}+3P^{(m)}\over
f^{\prime}}+\rho^{(c)}+3P^{(c)})\label{dy3},
\end{equation}
 where we have used the assumptions $N_{sur}={A\over
G}={4\pi\over GH^2}$ and $T={H\over2\pi}$ in $f(R)$ theory.
Moreover, due to the Bianchi identity, we get the continuity
equation as
\begin{equation}
{d\over dt}(\rho^{(c)}+{\rho^{(m)}\over f'})+3H({\rho^{(m)}+P^{(m)}\over f'}+\rho^{(c)}+P^{(c)})=0.
\end{equation}
Making use of the above continuity equation and integrating
Eq.(\ref{dy3}), we can finally obtain
 the other Friedmann equation as
\begin{equation}
H^2={8\pi G\over3}[{\rho^{(m)}\over f'}+\rho^{(c)}]\label{kim3}.
\end{equation}
Here we have also set the integration constant to vanish. The
above cosmological equations in f(R) theory are consistent with
the results in the literature\cite{Akbar:2006er}.\footnote{After we
completed this manuscript, we noticed the latest paper
\cite{Tu:2013gna} which overlaps with our content in this section,
although in \cite{Tu:2013gna} they employed a different area-entropy
relation such that the Friedmann dynamical equation is modified.
For us it seems not transparent to derive the corresponding
Friedmann kinematic equation from that modified dynamical
equation.} At the end of this section, we would like to present a
remark on the the area-entropy relation in scalar-tensor
gravity theory as well as in $f(R)$ theory. Once all the
contributions due to the scalar field or the curvature fluid are
absorbed into the total effective energy-momentum tensor, we may
think of them as some effective matter, such that the Einstein field
equations take the same form as the standard one. As a
result, the geometrical part in the field equation becomes the
usual Einstein tensor. Then, due to the purely geometrical nature
of the area-entropy relation, one would expect that it should be
preserved as usual, namely $S={A\over4G}$, though the
total effective energy-momentum tensor may provide important
corrections to the size and shape of the horizon.

\section{conclusions and discussions }
In this paper we have applied the idea of treating the cosmic space
as an emergent process to brane cosmology, scalar-tensor cosmology,
and f(R) gravity. We found that the corresponding cosmological equations in
these theories can be obtained such that the holographic nature of
this idea has been further examined in a more general setting.

 One of the key strategies employed in our paper is to define
the corresponding total effective energy-momentum tensor such that
the equation of motion in these theories can be written as the one
in standard general relativity. Such strategies are expected to be
applicable to other modified theories. For example, we may consider
a gravity theory with a Gauss-Bonnet invariant. We leave this for
further investigation.

We also expect this paradigm can be further improved to be
consistent with the thermodynamics of black holes when the
corrections due to the quantum effects of gravity are taken into
account. For instance, it is well known that such quantum gravity
effects may contribute a logarithmic correction to the entropy of
black holes. Correspondingly, it is reasonable to conjecture that
the number of degrees of freedom on the holographic screen is not
exactly proportional to one fourth of the area of the horizon,
but containing extra corrections like a logarithmic term. How to
take into account such extra contributions and consider their
impact on the modification of Friedmann equations should be an
interesting issue in the future.

\section*{Acknowledgements} We are grateful to Yu Tian
and Xiaoning Wu for helpful discussions. This work is partly
supported by the NSFC(Grants NO.11275208 and NO.11178002), the Jiangxi Young Scientists
(JingGang Star)Program and the 555 Talent Project of Jiangxi Province.


\begin{thebibliography}{99}

%\cite{Bardeen:1973gs}
\bibitem{Bardeen:1973gs}
  J.~M.~Bardeen, B.~Carter and S.~W.~Hawking,
  %``The Four laws of black hole mechanics,''
  Commun.\ Math.\ Phys.\  {\bf 31}, 161 (1973).  %%CITATION = CMPHA,31,161;%%


%\cite{Jacobson:1995ab}
\bibitem{Jacobson:1995ab}
  T.~Jacobson,
  %``Thermodynamics of space-time: The Einstein equation of state,''
  Phys.\ Rev.\ Lett.\  {\bf 75}, 1260 (1995)  [gr-qc/9504004].  %%CITATION = GR-QC/9504004;%%

  %\cite{Padmanabhan:2002sha}
\bibitem{Padmanabhan:2002sha}
  T.~Padmanabhan,
  %``Classical and quantum thermodynamics of horizons in spherically symmetric space-times,''
  Class.\ Quant.\ Grav.\  {\bf 19}, 5387 (2002)  [gr-qc/0204019].  %%CITATION = GR-QC/0204019;%%

  %\cite{Kothawala:2007em}
\bibitem{Kothawala:2007em}
  D.~Kothawala, S.~Sarkar, and T.~Padmanabhan,
  %``Einstein's equations as a thermodynamic identity: The Cases of stationary axisymmetric horizons and evolving spherically symmetric horizons,''
  Phys.\ Lett.\ B {\bf 652}, 338 (2007)  [gr-qc/0701002].  %%CITATION = GR-QC/0701002;%%
  %\cite{Cai:2005ra}
\bibitem{Cai:2005ra}
  R.~-G.~Cai and S.~P.~Kim,
  %``First law of thermodynamics and Friedmann equations of Friedmann-Robertson-Walker universe,''
  JHEP {\bf 0502}, 050 (2005)  [hep-th/0501055].  %%CITATION = HEP-TH/0501055;%%
  %\cite{Akbar:2006mq}
\bibitem{Akbar:2006mq}
  M.~Akbar and R.~-G.~Cai,
  %``Thermodynamic Behavior of Field Equations for f(R) Gravity,''
  Phys.\ Lett.\ B {\bf 648}, 243 (2007)  [gr-qc/0612089].  %%CITATION = GR-QC/0612089;%%
  %\cite{Cai:2008mh}
\bibitem{Cai:2008mh}
  R.~-G.~Cai, L.~-M.~Cao, Y.~-P.~Hu, and S.~P.~Kim,
  %``Generalized Vaidya Spacetime in Lovelock Gravity and Thermodynamics on Apparent Horizon,''
  Phys.\ Rev.\ D {\bf 78}, 124012 (2008)  [arXiv:0810.2610 [hep-th]].  %%CITATION = ARXIV:0810.2610;%%
  %\cite{Eling:2006aw}
\bibitem{Eling:2006aw}
  C.~Eling, R.~Guedens, and T.~Jacobson,
  %``Non-equilibrium thermodynamics of spacetime,''
  Phys.\ Rev.\ Lett.\  {\bf 96}, 121301 (2006)  [gr-qc/0602001].  %%CITATION = GR-QC/0602001;%%
  %\cite{Sheykhi:2007gi}
\bibitem{Sheykhi:2007gi}
  A.~Sheykhi, B.~Wang, and R.~-G.~Cai,
  %``Deep Connection Between Thermodynamics and Gravity in Gauss-Bonnet Braneworld,''
  Phys.\ Rev.\ D {\bf 76}, 023515 (2007)  [hep-th/0701261].  %%CITATION = HEP-TH/0701261;%%
  %\cite{Ge:2007yu}
\bibitem{Ge:2007yu}
  X.~-H.~Ge,
  %``First law of thermodynamics and Friedmann-like equations in braneworld cosmology,''
  Phys.\ Lett.\ B {\bf 651}, 49 (2007)  [hep-th/0703253].  %%CITATION = HEP-TH/0703253;%%
  %\cite{Gong:2007md}
\bibitem{Gong:2007md}
  Y.~Gong and A.~Wang,
  %``The Friedmann equations and thermodynamics of apparent horizons,''
  Phys.\ Rev.\ Lett.\  {\bf 99}, 211301 (2007)  [arXiv:0704.0793 [hep-th]].  %%CITATION = ARXIV:0704.0793;%%
  %\cite{Wu:2007se}
\bibitem{Wu:2007se}
  S.~-F.~Wu, B.~Wang, and G.~-H.~Yang,
  %``Thermodynamics on the apparent horizon in generalized gravity theories,''
  Nucl.\ Phys.\ B {\bf 799}, 330 (2008)  [arXiv:0711.1209 [hep-th]].  %%CITATION = ARXIV:0711.1209;%%

  %\cite{Ling:2009wj}
\bibitem{Ling:2009wj}
  Y.~Ling, W.~-J.~Li, and J.~-P.~Wu,
  %``Bouncing universe from a modified dispersion relation,''
  JCAP {\bf 0911}, 016 (2009)
  [arXiv:0909.4862 [gr-qc]].
  %%CITATION = ARXIV:0909.4862;%%

  %\cite{Guedens:2011dy}
\bibitem{Guedens:2011dy}
  R.~Guedens, T.~Jacobson, and S.~Sarkar,
  %``Horizon entropy and higher curvature equations of state,''
  Phys.\ Rev.\ D {\bf 85}, 064017 (2012)  [arXiv:1112.6215 [gr-qc]].  %%CITATION = ARXIV:1112.6215;%%

  %\cite{Sharif:2012zzd}
\bibitem{Sharif:2012zzd}
  M.~Sharif and M.~Zubair,
  %``Thermodynamics in f(R,T) Theory of Gravity,''
  JCAP {\bf 1203}, 028 (2012)  [arXiv:1204.0848 [gr-qc]].  %%CITATION = ARXIV:1204.0848;%%


  %\cite{Verlinde:2010hp}
\bibitem{Verlinde:2010hp}
  E.~P.~Verlinde,
  %``On the Origin of Gravity and the Laws of Newton,''
  JHEP {\bf 1104}, 029 (2011)  [arXiv:1001.0785 [hep-th]].  %%CITATION = ARXIV:1001.0785;%%



  %\cite{Padmanabhan:2009kr}
\bibitem{Padmanabhan:2009kr}
  T.~Padmanabhan,
  %``Equipartition of energy in the horizon degrees of freedom and the emergence of gravity,''
  Mod.\ Phys.\ Lett.\ A {\bf 25}, 1129 (2010)  [arXiv:0912.3165 [gr-qc]].  %%CITATION = ARXIV:0912.3165;%%


 %\cite{Padmanabhan:2003pk}
\bibitem{Padmanabhan:2003pk}
  T.~Padmanabhan,
  %``Gravitational entropy of static space-times and microscopic density of states,''
  Class.\ Quant.\ Grav.\  {\bf 21}, 4485 (2004)  [gr-qc/0308070].  %%CITATION = GR-QC/0308070;%%


%\cite{Padmanabhan:2012ik}
\bibitem{Padmanabhan:2012ik}
  T.~Padmanabhan,
  %``Emergence and Expansion of Cosmic Space as due to the Quest for Holographic Equipartition,''
  arXiv:1206.4916 [hep-th].  %%CITATION = ARXIV:1206.4916;%%

 %\cite{Cai:2012ip}
\bibitem{Cai:2012ip}
  R.~-G.~Cai,
  %``Emergence of Space and Spacetime Dynamics of Friedmann-Robertson-Walker Universe,''
  JHEP {\bf 1211}, 016 (2012)  [arXiv:1207.0622 [gr-qc]].  %%CITATION = ARXIV:1207.0622;%%

 %\cite{Yang:2012wn}
\bibitem{Yang:2012wn}
  K.~Yang, Y.~-X.~Liu, and Y.~-Q.~Wang,
  %``Emergence of Cosmic Space and the Generalized Holographic Equipartition,''
  Phys.\ Rev.\ D {\bf 86}, 104013 (2012)  [arXiv:1207.3515 [hep-th]].  %%CITATION = ARXIV:1207.3515;%%

 %\cite{Padmanabhan:2012gx}
\bibitem{Padmanabhan:2012gx}
  T.~Padmanabhan,
  %``Emergent perspective of Gravity and Dark Energy,''
  Res.\ Astron.\ Astrophys.\  {\bf 12}, 891 (2012)  [arXiv:1207.0505 [astro-ph.CO]].  %%CITATION = ARXIV:1207.0505;%%



%\cite{Ling:2010zc}
\bibitem{Ling:2010zc}
  Y.~Ling and J.~-P.~Wu,
  %``A note on entropic force and brane cosmology,''
  JCAP {\bf 1008}, 017 (2010)  [arXiv:1001.5324 [hep-th]].  %%CITATION = ARXIV:1001.5324;%%


%\cite{Akbar:2006er}
\bibitem{Akbar:2006er}
  M.~Akbar and R.~-G.~Cai,
  %``Friedmann equations of FRW universe in scalar-tensor gravity, f(R) gravity and first law of thermodynamics,''
  Phys.\ Lett.\ B {\bf 635}, 7 (2006)  [hep-th/0602156].  %%CITATION = HEP-TH/0602156;%%


%\cite{Capozziello:2005ku}
\bibitem{Capozziello:2005ku}
  S.~Capozziello, V.~F.~Cardone, and A.~Troisi,
  %``Reconciling dark energy models with f(R) theories,''
  Phys.\ Rev.\ D {\bf 71}, 043503 (2005)  [astro-ph/0501426].  %%CITATION = ASTRO-PH/0501426;%%

  %\cite{Nojiri:2006ri}
\bibitem{Nojiri:2006ri}
  S.~'i.~Nojiri and S.~D.~Odintsov,
  %``Introduction to modified gravity and gravitational alternative for dark energy,''
  eConf C {\bf 0602061}, 06 (2006)
  [Int.\ J.\ Geom.\ Meth.\ Mod.\ Phys.\  {\bf 04}, 115 (2007)]
  [hep-th/0601213].
  %%CITATION = HEP-TH/0601213;%%
  %1007 citations counted in INSPIRE as of 23 Jul 2013

  %\cite{Nojiri:2010wj}
\bibitem{Nojiri:2010wj}
  S.~'i.~Nojiri and S.~D.~Odintsov,
  %``Unified cosmic history in modified gravity: from F(R) theory to Lorentz non-invariant models,''
  Phys.\ Rept.\  {\bf 505}, 59 (2011)
  [arXiv:1011.0544 [gr-qc]].
  %%CITATION = ARXIV:1011.0544;%%
  %423 citations counted in INSPIRE as of 23 Jul 2013

  %\cite{DeFelice:2010aj}
\bibitem{DeFelice:2010aj}
  A.~De Felice and S.~Tsujikawa,
  %``f(R) theories,''
  Living Rev.\ Rel.\  {\bf 13}, 3 (2010)  [arXiv:1002.4928 [gr-qc]].  %%CITATION = ARXIV:1002.4928;%%

%\cite{Sotiriou:2008rp}
\bibitem{Sotiriou:2008rp}
  T.~P.~Sotiriou and V.~Faraoni,
  %``f(R) Theories Of Gravity,''
  Rev.\ Mod.\ Phys.\  {\bf 82}, 451 (2010)  [arXiv:0805.1726 [gr-qc]].  %%CITATION = ARXIV:0805.1726;%%


%\cite{Jaime:2012gc}
\bibitem{Jaime:2012gc}
  L.~G.~Jaime, L.~Patino, and M.~Salgado,
  %``f(R) Cosmology revisited,''
  arXiv:1206.1642 [gr-qc].  %%CITATION = ARXIV:1206.1642;%%


%\cite{Bamba:2011jq}
\bibitem{Bamba:2011jq}
  K.~Bamba, C.~-Q.~Geng, and S.~Tsujikawa,
  %``Thermodynamics in Modified Gravity Theories,''
  Int.\ J.\ Mod.\ Phys.\ D {\bf 20}, 1363 (2011)  [arXiv:1101.3628 [gr-qc]].  %%CITATION = ARXIV:1101.3628;%%

%\cite{Tu:2013gna}
\bibitem{Tu:2013gna}
 F.~-Q.~Tu and Y.~-X.~Chen,
  %``Emergence of spaces and the dynamic equations of FRW universes in the $f(R)$ theory and deformed Ho\v{r}ava-Lifshitz theory,''
  JCAP {\bf 1305}, 024 (2013)
  [arXiv:1303.5813 [hep-th]].
  %%CITATION = ARXIV:1303.5813;%%
  %2 citations counted in INSPIRE as of 23 Jul 2013










%\cite{Chimento:2003ie}

\end{thebibliography}
\end{document}